\documentclass[twocolumn,trackchanges]{aastex62}

\usepackage{amsmath}
\graphicspath{{./}{figures/}}
\usepackage{graphicx}
\usepackage{booktabs}
\usepackage{comment}

\usepackage{savesym}
\savesymbol{tablenum}
\usepackage{siunitx}
\restoresymbol{SIX}{tablenum}

\shorttitle{ }
\shortauthors{You et al.}

\begin{document}
\title{Standard-siren cosmology using gravitational waves from binary black holes}

\email{you\_zhiqiang@whu.edu.cn\\
zhuxingjiang@gmail.com\\
zhuzh@whu.edu.cn}

\author{Zhi-Qiang You}
\affiliation{School of Physics and Technology, Wuhan University, Wuhan 430072, China}
\affiliation{School of Physics and Astronomy, Monash University, Clayton, Vic 3800, Australia}
\affiliation{OzGrav: The ARC Centre of Excellence for Gravitational Wave Discovery, Clayton, Vic 3800, Australia}

\author[0000-0001-7049-6468]{Xing-Jiang Zhu}
\affiliation{School of Physics and Astronomy, Monash University, Clayton, Vic 3800, Australia}
\affiliation{OzGrav: The ARC Centre of Excellence for Gravitational Wave Discovery, Clayton, Vic 3800, Australia}

\author[0000-0001-7288-2231]{Gregory Ashton}
\affiliation{School of Physics and Astronomy, Monash University, Clayton, Vic 3800, Australia}
\affiliation{OzGrav: The ARC Centre of Excellence for Gravitational Wave Discovery, Clayton, Vic 3800, Australia}

\author[0000-0002-4418-3895]{Eric Thrane}
\affiliation{School of Physics and Astronomy, Monash University, Clayton, Vic 3800, Australia}
\affiliation{OzGrav: The ARC Centre of Excellence for Gravitational Wave Discovery, Clayton, Vic 3800, Australia}

\author{Zong-Hong Zhu}
\affiliation{School of Physics and Technology, Wuhan University, Wuhan 430072, China}
\affiliation{Department of Astronomy, Beijing Normal University, Beijing 100875, China}
\newcommand{\red}[1]{\textcolor{red}{#1}}
\begin{abstract}
Gravitational-wave astronomy provides a unique new way to study the expansion history of the Universe.
In this work, we investigate the impact future gravitational-wave observatories will have on cosmology. Third-generation observatories like the Einstein Telescope and Cosmic Explorer will be sensitive to essentially all of the binary black hole coalescence events in the Universe. Recent work by \cite{farr2019future} points out that features in the stellar-mass black hole population break the mass-redshift degeneracy, facilitating precise determination of the Hubble parameter without electromagnetic counterparts or host galaxy catalogues. 
Using a hierarchical Bayesian inference model, we show that with one year of observations by the Einstein Telescope, the Hubble constant will be measured to $\lesssim 1\%$.
We also show that this method can be used to perform Bayesian model selection between cosmological models.
As an illustrative example, we find that a decisive statement can be made comparing the $\Lambda$CDM and RHCT cosmological models using two weeks of data from the Einstein Telescope.
\end{abstract}


\section{Introduction}
The first direct detection of gravitational waves (GWs) by LIGO \citep{2015CQGra..32g4001L} and Virgo \citep{2015CQGra..32b4001A} in 2015 \citep{2016PhRvL.116f1102A} opened a new window for the study of our Universe \citep{abbott2017gw170814,abbott2016gw151226,2016PhRvL.116f1102A}. 
Since then, dozens more GW events and candidates\footnote{https://gracedb.ligo.org/superevents/public/O3/} have been reported; see \cite{abbott2019gwtc} and \citet{GWTC-2} for a catalogue of 50 events published by LIGO/Virgo so far.
The first GW signal from a binary neutron star merger, GW170817~\citep{abbott2017gw170817}, was accompanied by a counterpart detected across the electromagnetic (EM) spectrum~\citep{2017ApJ...848L..12A}.
Combining the luminosity distance determined from GW observations with the redshift inferred from EM data, the Hubble constant was measured to be $H_0=70.0^{+12.0}_{-8.0}\, \si{km \ s^{-1} Mpc^{-1}}$ \citep{ligo2017gravitational}.
Subsequent observations of the radio counterpart of GW170817 using very long baseline interferometry broke the luminosity distance-viewing angle degeneracy, which improved the measurement to $ H_0=  68.9^{+4.7}_{-4.6} \,\si{km \ s^{-1}Mpc^{-1}}$ \citep{hotokezaka2019hubble}.
Gravitational-wave observations of compact binary coalescences are ``standard sirens'' \citep{schutz1986determining} because they provide an independent way of constraining the expansion history of the Universe, complementary to other cosmological probes, including supernovae \citep{riess20113,riess2009redetermination}, the cosmic microwave background \citep{adam2016planck,komatsu2011seven,lewis2002cosmological}, baryon acoustic oscillations \citep{beutler20116df,percival2010baryon}, and gravitational lensing \citep{schrabback2010evidence,liao2017precision}.

While multi-messenger observations (GW + EM) provide a powerful tool for cosmology, there are challenges with this approach.
First, compact binary detections are dominated by binary black holes (BBHs), for which no EM counterparts are expected.
Second, the GW sky localization is often large ($> 10^2$ deg$^2$), at least for the current network of detectors, making EM follow-ups difficult \citep[see, e.g.,][for the case of GW190425, the second binary neutron star merger]{abbott2020gw190425}.
Finally, multi-messenger cosmology will eventually be limited by incomplete galaxy catalogues up to the redshifts observable by third-generation detectors.

In this work, we explore an alternative approach for performing GW cosmology without using EM counterparts or host galaxy catalogues. 
We make use of a unique feature in the black hole mass distribution: a mass gap between $\sim 50 \, \si{M_{\odot}}$ and $\sim 150 \, \si{M_{\odot}}$.
Such a mass gap is thought to exist due to the pair instability supernova (PISN) process \citep{fowler1964neutrino,heger2010nucleosynthesis,belczynski2016effect} and has found support\footnote{After this paper was submitted, \citet{abbott2020gw190521} reported the discovery of GW190521 -- a BBH merger with component masses within the PISN mass gap. Subsequent analysis found that its component masses could fall outside the gap under certain priors \citep{Maya190521,Nitz190521}. It has also been suggested that GW190521 might involve one or two second-generation BHs formed dynamically in dense stellar environments \citep{GW190521prop,Romero-Shaw190521}.} in the observed population of LIGO/Virgo BBHs \citep{abbott2019gwtc}.
Recently, \cite{farr2019future} considered this feature and showed that it is possible to measure the Hubble constant to a precision of 6\% using BBHs detected with the advanced (i.e., second-generation) detector network after one year of operation at design sensitivity.

Here we investigate the capability of proposed third-generation detectors, such as the Einstein Telescope (ET) \citep{2010CQGra..27s4002P} and Cosmic Explorer (CE) \citep{abbott2017exploring}.
These detectors, expected to be operational in the 2030s, will be able to detect BBH mergers throughout the Universe, yielding $\sim 10^4 - 10^7$ discoveries per year \citep{2011arXiv1108.1423S}.
The high-redshift reach of ET/CE ($z\gtrsim 6$) complements the supernova standard-candle observations, which are limited to relatively low redshifts \citep[$z\lesssim 2.4$, e.g.,][]{graur2014type}.
This may help settle the ``Hubble tension'' found among different measurements of $H_0$ \citep[e.g.,][]{adam2016planck}.
As a proof of principle, we also demonstrate cosmological model selection by comparing the standard $\Lambda$CDM model with the RHCT model \citep{melia2009cosmological,melia2012r}.

This paper is organized as follows. In Section \ref{sec:BBHcatalog} we describe our model of BBH mass distribution and simulate a population of BBH events representative of third-generation detectors.
In Section \ref{sec:Bayes}, we introduce the Bayesian hierarchical inference method used for the analysis.
In Section \ref{sec:results}, we present results of comological parameter estimation and model selection.
In Section \ref{sec:conclusions}, we summarize this work.

\section{The simulated BBH population}
\label{sec:BBHcatalog}
\subsection{Black hole mass distribution}
\label{sec:massDist}
{We model the black hole mass distribution in the source frame using the ``{\tt{POWER LAW + PEAK}}'' model from the LIGO--Virgo second gravitational-wave transient catalog (GWTC-2) population analysis \citep[see Appendix B.2 in][]{GWTC-2pop}. It consists of a power-law distribution and a Gaussian peak that represents the build-up of black holes due to pulsational PISNe.}
In this model, the probability distribution of the primary black hole mass ($m_1$) is given by:
\begin{equation} \label{pm1}
P({m_1}) = [(1-\lambda)P_{\rm{pow}}({m_1}) + \lambda P_{\rm{pp}}({m_1})]S(m_1,m_{\text{min}},\delta_m)\, ,
\end{equation}
where $\lambda$ is a mixing fraction parameter that gives the weight of the Gaussian component, and $S$ is a smoothing function which rises from 0 at $m_{\rm{min}}$ to 1 at $m_{\rm{min}}+\delta m$ \citep[see Eq. (B6) in][]{GWTC-2pop}.

The power-law distribution is given by:
\begin{equation}
P_{\rm{pow}}({m_1})  \propto {({m_1})}^{-\alpha} \mathcal{H}(m_{\rm{max}}-{m_1})\, ,
\end{equation}
where $\alpha$ is the power-law index, $\mathcal{H}$ is the Heaviside step function, and $m_{\rm{min}}$ ($m_{\rm{max}}$) is the minimum (maximum) black-hole mass.
The Gaussian component, with a mean $m_{\rm{pp}}$ and a standard deviation $\sigma_{\rm{pp}}$, is given by:
\begin{equation}
P_{\rm{pp}}({m_1}) \propto \exp\left[-\frac{({m_1} - m_{\rm{pp}})^2}{2\sigma_{\rm{pp}}^2}\right] \, .
\end{equation}
Assuming a power-law distribution (with index $\beta$) of mass ratio ($m_{2}/m_{1}\leq 1$), the conditional probability of the secondary black hole mass ($m_2$) given $m_1$ can be written as:
\begin{equation}\label{pm2}
P({m_2}|m_1)\propto\left( {\frac{{m_2}}{{m_1}}}\right )^{\beta}S({m_2},m_{\rm{min}},\delta m). 
\end{equation}

In this work, we assume that the intrinsic BBH mass distribution does not evolve over cosmic time.
{Based on the analysis of \citet{GWTC-2pop}, we adopt the following fiducial model parameters: a Gaussian peak at $m_{\text{pp}}=33.5\,  \si{M_{\odot}}$, a high-mass cut-off at $m_{\text{max}}=65\,  \si{M_{\odot}}$, mass ratio index $\beta=1.3$, primary mass power-law index $\alpha=2.5$, Gaussian peak weight $\lambda=0.1$, minimum mass $m_{\text{min}}=5\, \si{M_{\odot}}$, Gaussian peak width $\sigma_{\rm{\rm{pp}}}=5\, \si{M_{\odot}}$, and low-mass turn-on  $\delta_m=4\, \si{M_{\odot}}$.} To reduce the computational cost, we choose to hold $m_{\text{max}}$ and $m_{\text{min}}$ fixed at injection values for this study.
We discuss potential impacts on the results from the above assumptions in section \ref{sec:discuss}.

\subsection{Generation of mock BBH catalogues}
We assume a flat Friedmann-Robertson-Walker Universe, and use the $\Lambda$CDM model as the fiducial model throughout this paper. The luminosity distance can be written as:
\begin{equation}
    D_L=(1+z){\int^z_0}c/H(z')dz'\, ,
\end{equation}
where $H(z)=H_0{\sqrt{(1+z)^3\Omega_m+\Omega_{\Lambda}}}$ is the Hubble parameter given a
dark energy equation of state of $w=-1$, $\Omega_m$ is the matter density and $\Omega_{\Lambda}=1-\Omega_m$ is the dark energy density. The comoving distance $r$ is given by $r(z) = D_{L}/(1+z)$.

For illustrative purposes, we also consider the RHCT cosmology \citep{melia2012r}, where $r(z) H_{0} =c \ln(1+z)$. We note that \citet{bilicki2012we} have shown that the RHCT model is inconsistent with observations of supernovae at low redshifts $(z \lesssim 0.5)$. Nevertheless, we use the RHCT model in this work as an example for cosmological model selection.

The number distribution of BBH events in the parameter space of $(D_{L}, m_{1}, m_{2})$ is 
\begin{equation} 
\label{eq:dNdDLdm}
   \frac{dN}{dD_L dm_1 d m_2}=\frac{d z}{d D_L}R(z)\frac{dV_c}{dz}\frac{T_{\rm{obs}}}{1+z}P(m_1,m_2)\, ,
\end{equation}
where $N$ is the number of events collected within the observation time $T_{\rm{obs}}$.
Here, the comoving volume is $dV_c/dz=4\pi c r^{2}(z)/H(z)$.
$R(z)$ is the BBH merger rate density as a function of redshift.
The local rate density, $R(z=0)$, was estimated to be $24\, \si{{{{Gpc}}^{-3}{{yr}}^{-1}}}$ \citep{GWTC-2pop}. 
The merger rate is a convolution of the binary formation rate with the distribution of the time delays \citep{nakar2007short,Zhu11bbh-gwb}.
We assume the binary formation rate closely follow the cosmic star formation rate, for which we take the model in \cite{madau2014cosmic}.
However, to demonstrate the influence of star formation rate uncertainty on the inference of cosmological parameters, we also consider an alternative model by \citet{2012ApJ...744...95R}.
We assume the time delay between binary formation and binary merger follows a power-law distribution $P(t_{d})\propto (t_{d})^{\zeta}$ with a minimum delay time of $t_{d}^{\rm{min}}$.
We treat $\zeta$ and $t_d^\text{min}$ as free parameters so that we can investigate how uncertainty in the delay time distribution affects our measurement of cosmological parameters.

The measured masses in the detector frame are related to source-frame masses by:
\begin{equation} 
\label{eq:mz}
 m_{1,2}^z={m_{1,2}}\big[1+z_{1,2}(D_{L})\big]\, .
\end{equation}
The number distribution of BBH events given in Equation (\ref{eq:dNdDLdm}) is conditional upon a collection of hyperparameters, including cosmological parameters $H_0$ and $\Omega_m$, black hole mass distribution parameters introduced in Section \ref{sec:massDist}, and parameters that determine the cosmic star formation rate and delay time distribution.

In this proof-of-principle study, we are mostly concerned with the cosmological parameters.
For star formation rate,  we take the parameterized form given by Eq. 15 of \citet{madau2014cosmic}, and adopt the following parameter values $b=2.7,\, c=2.9,\, d=5.6$.
For the delay time distribution, we use $\zeta=-1$ and $t_{d}^{\rm{min}}=50\, \si{Myr}$. To generate mock BBH catalogues, we set $H_0 = 70\, \si{km \ s^{-1}Mpc^{-1}}$, $\Omega_m=0.3$.
Integrating Equation (\ref{eq:dNdDLdm}) over masses and luminosity distance, we find that $\sim 10^{5}$ BBH events will be detected within one year by ET under our fiducial model.

\begin{figure}
 \centering
 \includegraphics[width=3.33in]{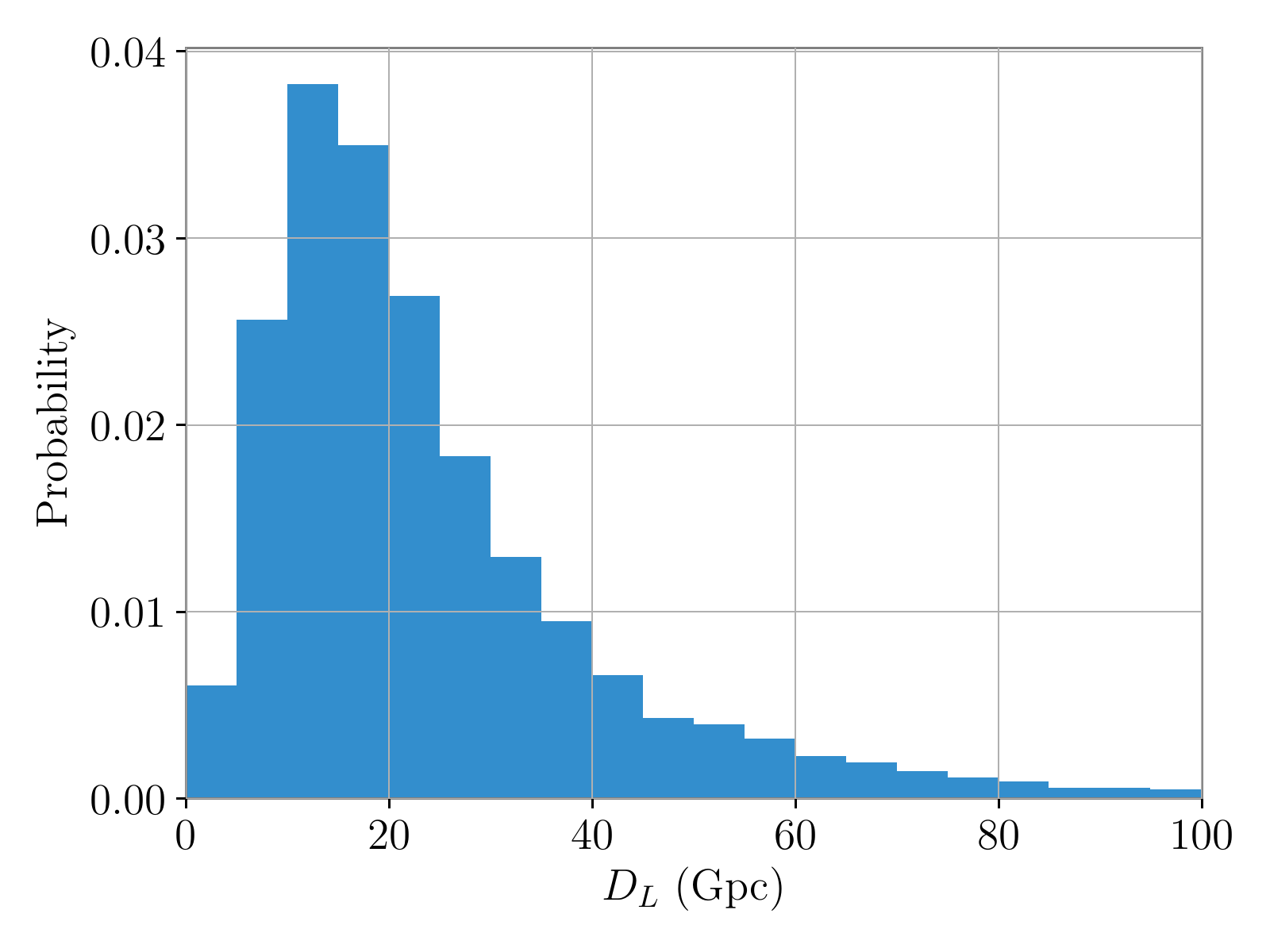}
 \includegraphics[width=3.13in]{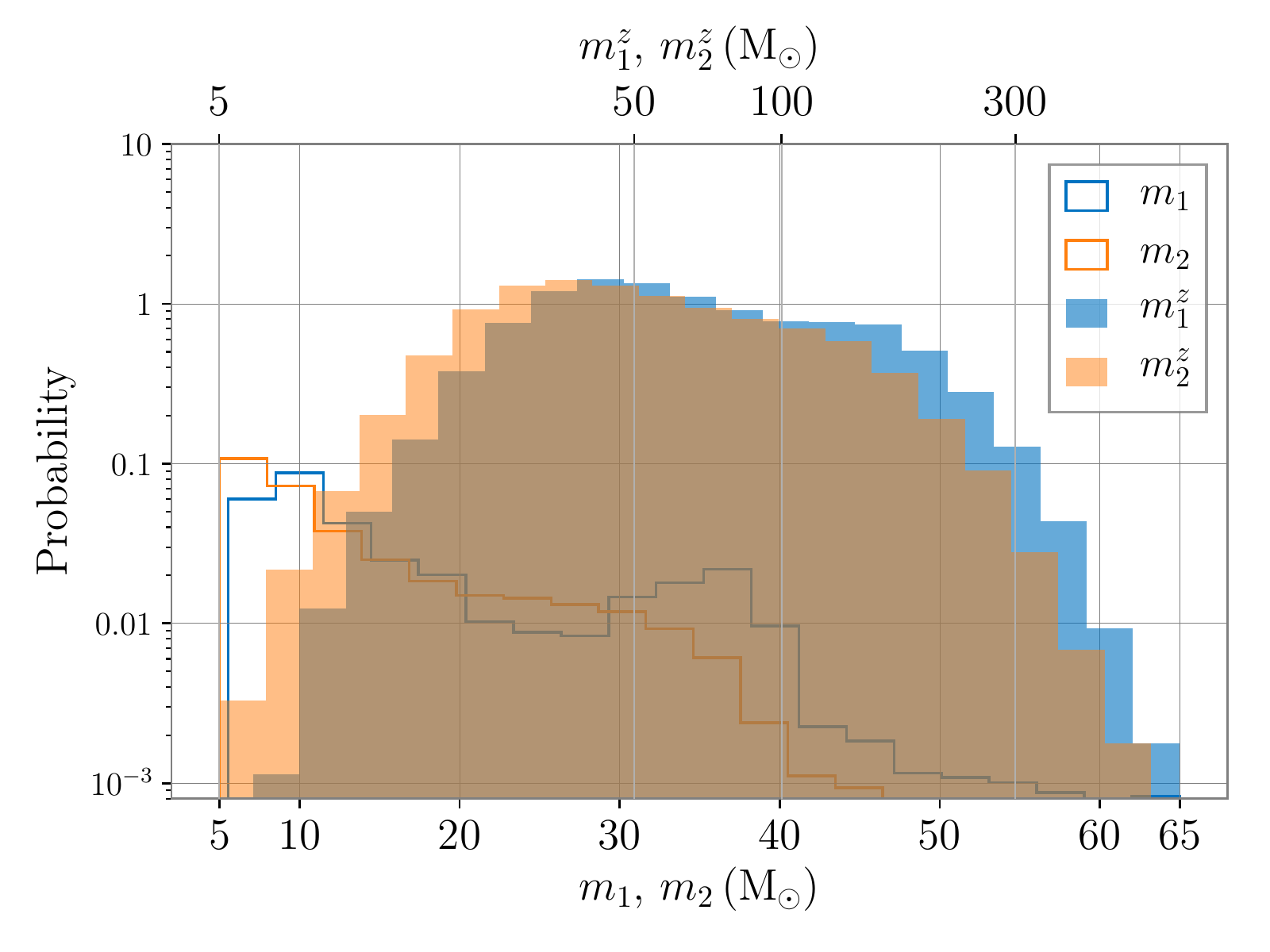}
 \caption{The distribution of luminosity distance ($D_L$) and black hole masses ($m_1, m_2$) for a simulated BBH population detectable by third-generation detectors, where the probability is normalized with the logarithm of mass.}
 \label{fig:dlm1m2}
\end{figure}

Figure \ref{fig:dlm1m2} shows the distribution of luminosity distance $D_L$ (top panel), and black hole mass (bottom panel) in the source frame $(m_1, m_2)$ and in the detector frame $(m_1^z, m_2^z)$ for our simulated BBH population. We apply an upper limit of $D_L$ at $100\,\si{Gpc}$ , corresponding to a redshift of $\sim 10$, beyond which the number of detectable BBHs is likely negligible.
The luminosity distance distribution peaks at around $10 \,\si{Gpc}$ ($z\sim 1.5$), as expected from cosmic star formation rate.
Because the population is dominated by relatively high-redshift events ($z\gtrsim 2$), the distribution of lab-frame masses is much smoother than that of source-frame masses.

We simulate up to $10^5$ BBH events following the distributions illustrated in Figure \ref{fig:dlm1m2}, with other source parameters drawn from their respective standard default distributions \citep[e.g., Table 1 of][]{ashton2019bilby}.
Ideally we would add these BBH signals to Gaussian noise realizations generated from a detector sensitivity curve, then perform Bayesian inference using software packages like {\tt BILBY} \citep{ashton2019bilby}, and obtain posterior distribution of source parameters. Since only distributions of luminosity distance and black hole masses contain information about cosmology, the posterior distributions are marginalized over parameters other than $(D_L, m_1^z, m_2^z)$.
These posterior distributions for individual BBH events are combined in a hierarchical Bayesian framework to estimate hyperparameters $H_0$ and $\Omega_m$, as we will describe in the next section.

\begin{figure}[htp]
 \centering
 \includegraphics[width=3.3in]{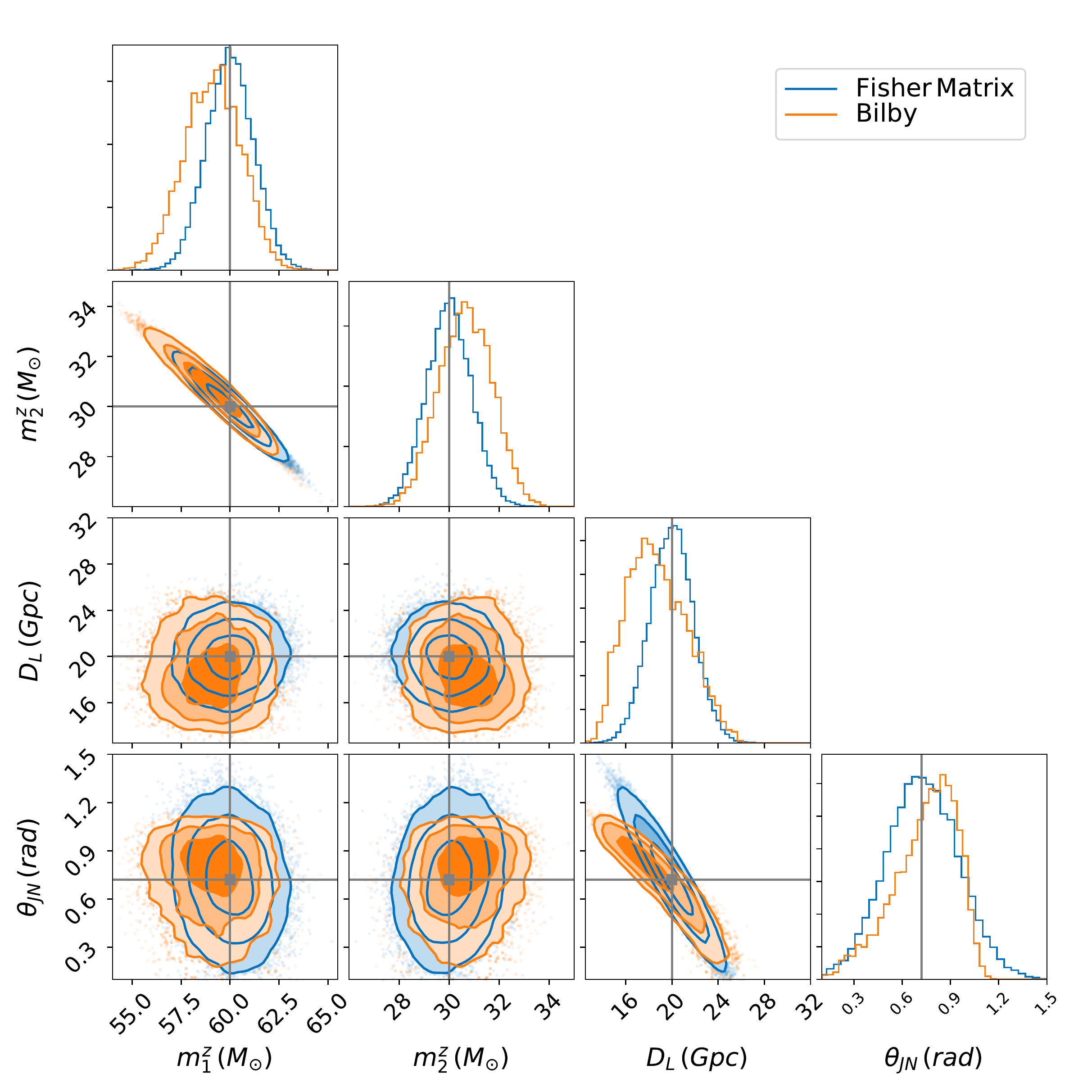}
 \caption{The joint posterior distribution of detector-frame masses $m_1^z$, $m_2^z$, $D_L$, $\theta_{JN}$ for a simulated BBH event obtained using the full parameter-estimation code {\tt BILBY} (orange), the Fisher Matrix approximation (blue).
The solid lines mark the injection values: $m^z_1=60 \, \si{M_\odot}$, $m^z_2=30\, \si{M_\odot}$, $D_L=20\, \si{Gpc}$, $\Theta_{JN}=0.72$ rad. The signal-to-noise ratio of this injection is 15.4.}
 \label{fig:fmmcmc}
\end{figure}

In practice, running full parameter estimation for $10^5$ events is computationally challenging.
Therefore, we employ the Fisher Information Matrix approach \citep{vallisneri2008use,rodriguez2013inadequacies} to construct the posterior distributions for individual BBH events.
The known correlations between $m_1^z$ and $m_2^z$, and between $D_L$ and inclination angle $\theta_{JN}$, are properly accounted for in our analysis.
Figure \ref{fig:fmmcmc} compares the joint posterior distribution of $(m_1,m_2,D_L,\theta_{JN})$ derived by Fisher Matrix and that returned from a full parameter estimation with {\tt BILBY}, for one BBH event. 
The posterior distribution widths from both methods are similar, which demonstrates the effectiveness of our approach.

\section{Bayesian inference}

\label{sec:Bayes}
In this section, we employ hierarchical Bayesian inference to compute posterior distributions of hyperparameters that describe the BBH population.
Extensive descriptions of Bayesian hierarchical inference can be found in \citet{thrane2019introduction}. We introduce it briefly below.

We introduce the conditional prior, i.e., the prior distribution of BBH source parameters $\theta$ conditional upon some hyperparameters $\Lambda$. For this work, we are concerned with the prior distribution of $(D_L$, ${m_1}$, ${m_2}) \in \theta$ that is dependent on cosmological hyperparameters $(H_0, \Omega_m) \in \Lambda$ via Equation (\ref{eq:dNdDLdm}). The conditional priors is denoted as $\pi(\theta|\Lambda)$.

The hyper-likelihood ${\cal L}(h|\Lambda)$ is related to the regular likelihood ${\cal L}(h|\theta)$ by:
\begin{equation} \label{hylike1}
{\cal L}(h|\Lambda)=\int {\cal L}(h|\theta)\pi(\theta|\Lambda)d\theta \, ,
\end{equation}
where $h$ denotes the gravitational-wave data. In hierarchical inference, we have access to the posterior distribution of parameters of individual BBHs $P(\theta|h)$, which is related to the regular likelihood through the Bayes' theorem
\begin{equation} \label{regularlike}
{\cal L}(h|\theta)={\cal Z}(h)\frac{P(\theta|h)}{\pi(\theta)}
\, ,
\end{equation}
where ${\cal Z}(h)$ is the evidence and $\pi(\theta)$ is prior used for the parameter estimation of individual events.
Rewriting Equation (\ref{hylike1}) by replacing the integral with the summation over discrete posterior samples \citep[e.g.,][]{hogg2018data}, we obtain
\begin{equation} \label{hylike2}
{\cal L}(h|\Lambda)=\frac{{\cal Z}(h)}{n}\sum^{n}_{k=1} \frac{\pi(\theta^{k}|\Lambda)}{\pi(\theta^{k})}\, ,
\end{equation}
where $n$ is the number of posterior samples for individual events.
Combining $N$ independent events, we obtain the total likelihood
\begin{equation} \label{hylike}
{\cal L}_{\rm tot}(\textit{\textbf{h}}|\Lambda)=\prod^N_{i=1} \frac{{\cal Z}(h_i)}{n_i}\sum^{n_i}_{k=1} \frac{\pi(\theta^k_i|\Lambda)}{\pi(\theta^k_i)}\, ,
\end{equation}
where $\textit{\textbf{h}}$ denotes the collection of data $\{h_i\}$.
The hyper-posteriors are given by $P(\Lambda|\textit{\textbf{h}}) \propto {\cal L}_{\rm tot}(\textit{\textbf{h}}|\Lambda) \pi(\Lambda)$, with $\pi(\Lambda)$ being the prior distribution of hyperparameters.

In order to perform model selection, it is also necessary to calculate the hyper-evidence given a model $M$
\begin{equation}
{\cal{Z}}_{\rm tot}( \textit{\textbf{h}}|M)=\int {\cal{L}}_{\rm tot}(\textit{\textbf{h}}|\Lambda,M)\pi (\Lambda|M)d\Lambda\, .
\label{eqn:evidence}
\end{equation}
The Bayes factor (BF) between model $M_1$ and model $M_2$ is
\begin{equation} \label{BF}
{\rm BF}^{1}_{2}=\frac{{\cal{Z}}_{\rm tot}( \textit{\textbf{h}}|M_{1})}{{\cal{Z}}_{\rm tot}( \textit{\textbf{h}}|M_{2})}\, .
\end{equation}
We impose a threshold of the natural logarithm of BF at $\ln(BF)=8$ as the point when one model is significantly favoured against another \citep[e.g.,][]{2003itil.book.....M}.

\section{Results}
\label{sec:results}
In this section, we present results of hyperparameter estimation and cosmological model selection using the simulated BBH population that is expected to be detected by third-generation detectors such as ET.

\subsection{Hyperparameter estimation}
\label{sec:hyperpe}
\begin{figure}
 \centering
 \includegraphics[width=3.3in]{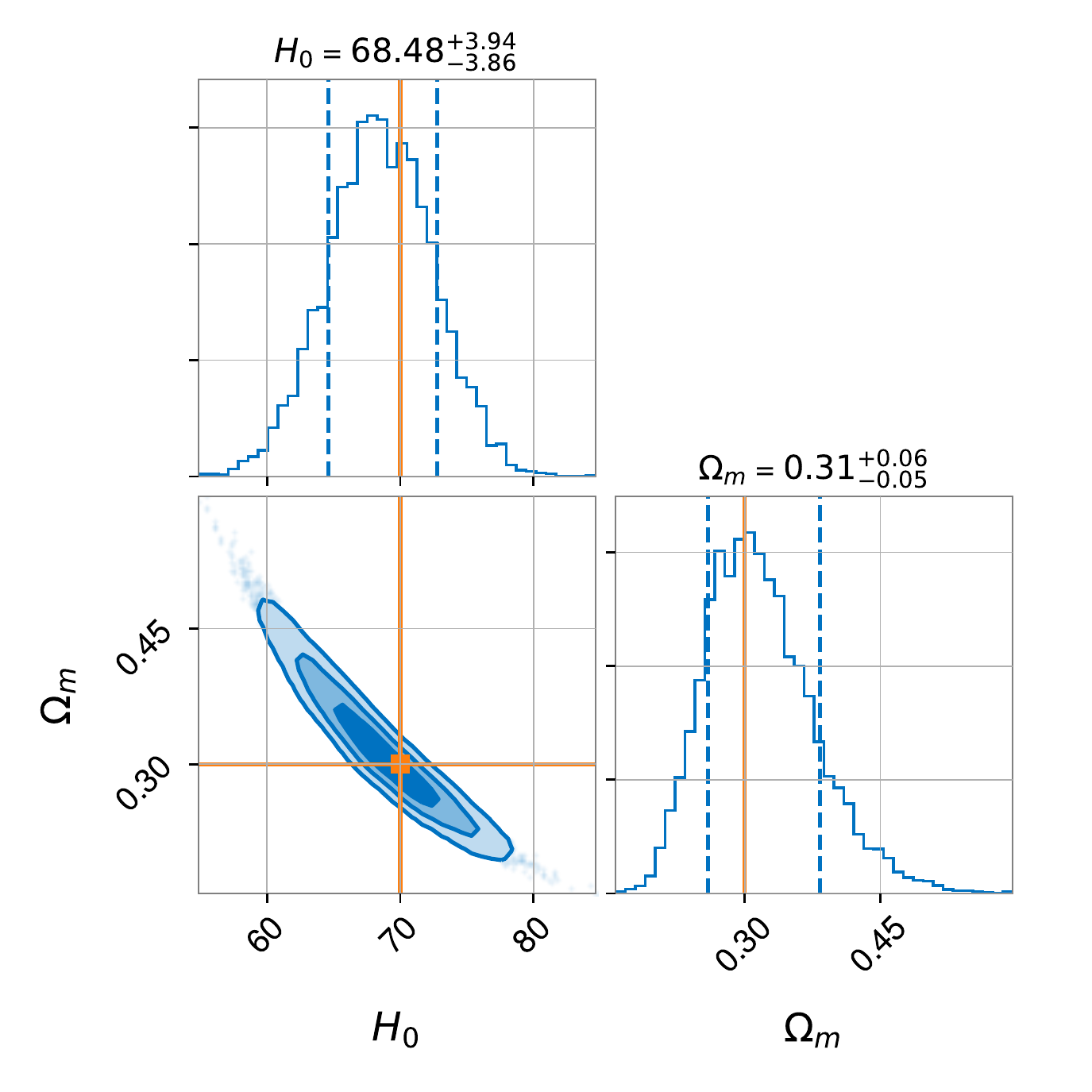}
 \caption{Joint posterior distribution of the Hubble constant ($H_0$) and matter density ($\Omega_m$) in the $\Lambda$CDM model estimated using $10^{3}$ BBH events. The 2-D contour regions denote the 1-$\sigma$, 2-$\sigma$ and 3-$\sigma$ credible regions and the orange lines indicate the true values.}
 \label{fig:hypos}
\end{figure}

\begin{figure}
 \centering
 \includegraphics[width=3.3in]{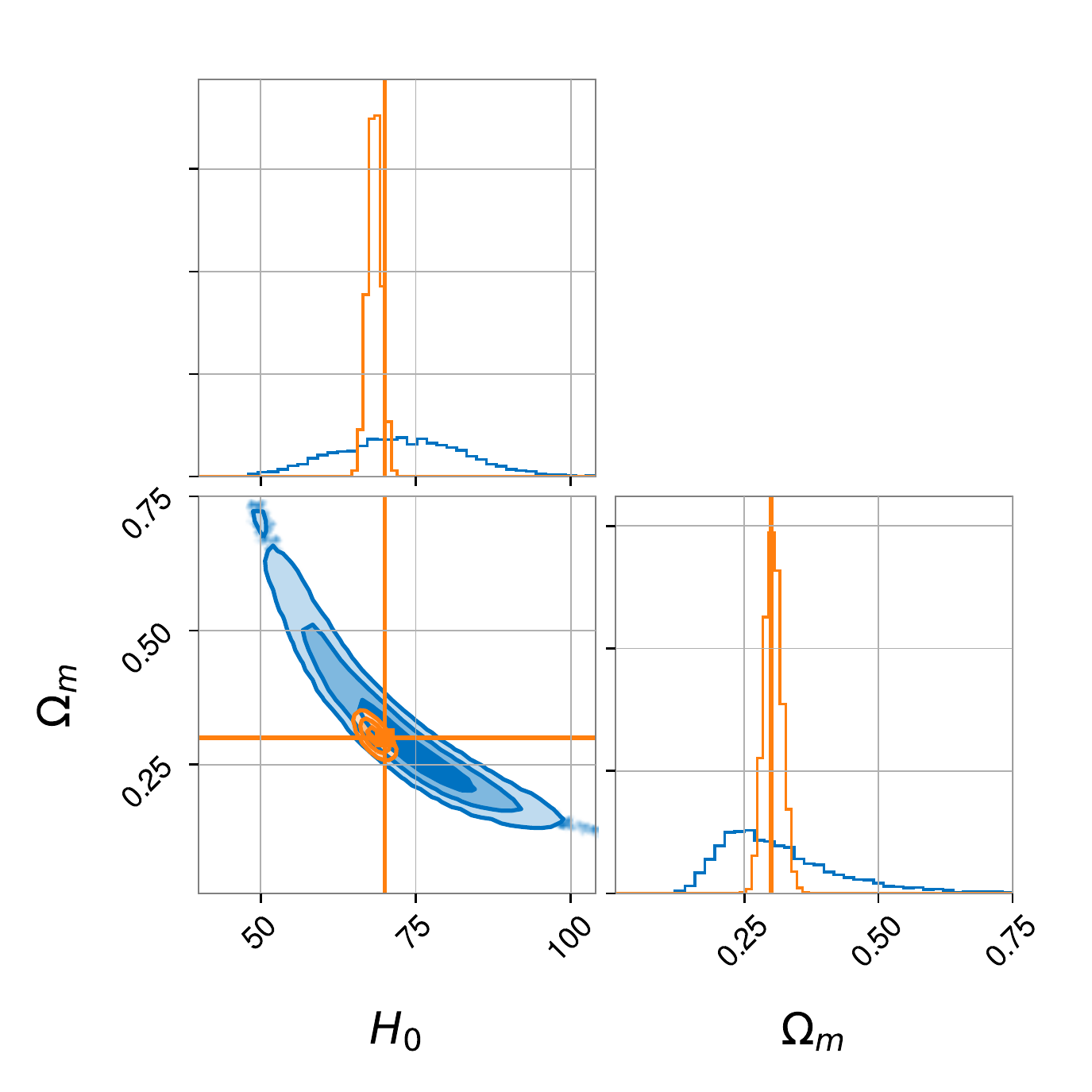}
 \caption{Posteriors distribution for $(H_{0}, \Omega_{m})$ with $10^{3}$ zeros-error injections. Blue contours are obtained by marginalizing over uncertainties in other hyperparameters $(\zeta,\,t_{d}^{\rm{min}},\,b,\,c,\,d,\,\delta_m,\,\alpha,\,m_{\rm{pp}},\,\delta_{\rm{pp}},\,\lambda,\,\beta)$, whereas the orange is reconstructed with non-cosmological parameters fixed at injection values.}
 \label{fig:2p8p}
\end{figure}

Figure \ref{fig:hypos} shows the joint posterior distribution of $H_0$ and $\Omega_m$ using $10^3$ BBH events detected with ET, while ignoring the delay time between binary formation and binary merger and assuming we know the cosmic star formation rate and black hole mass distribution \textit{a priori}. In our analysis, uniform priors are used: $H_0 \in [40,105] \, \si{ km \ s^{-1}Mpc^{-1}}$ and $\Omega_m \in [0,0.75]$.
In this example, the Hubble constant is measured with a precision of $5.6\%$.
By performing this analysis for a range of $N$ (the number of BBH events) assuming zero measurement uncertainty of luminosity distance and black hole masses (which we call zero-error injections), we find the measurement precision of $H_0$ scales linearly with $\sqrt{N}$. We expect that one year operation of ET, yielding $\sim 10^5$ BBH detections, will allow $H_0$ to be measured within $\approx 0.6\%$.
However, this result is too optimistic as it does not account for uncertainties in cosmic star formation rate, delay time distribution and black hole mass distribution.
In Figure \ref{fig:sfrd} of the Appendix, we show that the estimates of $(H_{0}, \Omega_{m})$ are biased if an incorrect model of star formation rate is used.

To demonstrate how the marginalization over unknowns in non-cosmological parameters affects our ability to measure $H_{0}$ and $\Omega_{m}$, we repeat the analysis using $10^{3}$ zero-error injections.
In the reference case, only $H_{0}$ and $\Omega_{m}$ are considered as free parameters.
We find that the $1-\sigma$ credible interval of $H_0$ is respectively increased by 35\%, 50\%, and 250\% if we add delay time distribution parameters, cosmic star formation rate parameters and black hole mass distribution parameters, respectively.
Figure \ref{fig:2p8p} compares the posterior distribution of $(H_{0},\Omega_{m})$ for the reference case (orange) and the case where non-cosmological parameters are sampled and marginalized over (blue).
Full posteriors (Figure \ref{fig:8p}), along with their priors, of all hyperparameters in the latter case are presented in the Appendix.
We find the marginalization over uncertainties in non-cosmological hyperparameters reduce the measurement precision of $(H_{0},\Omega_{m})$ by about an order of magnitude.
Therefore, we conclude that one year observation of ET will constrain the Hubble constant to a few percent given our current knowledge of the black hole mass distribution, the cosmic star formation rate, and the binary merger delay time distribution.
If/when our understanding of the above quantities is improved, which is plausible in the ET era, a sub-percent measurement precision is likely.

\subsection{Model selection}

\begin{figure}
 \centering
 \includegraphics[width=3.3in]{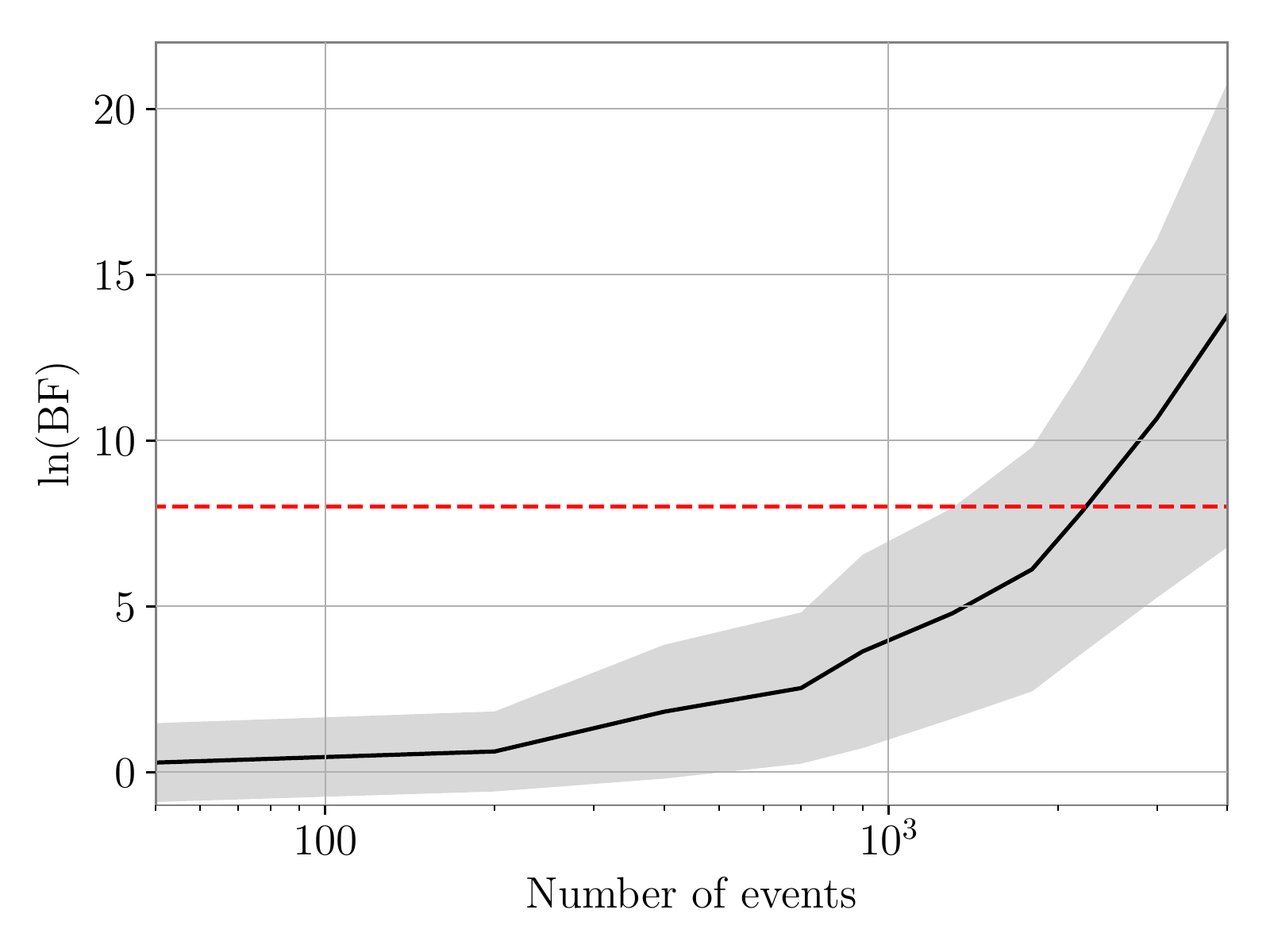}
 \caption{The natural logarithm of Bayes Factor ln(BF) as a function of the number of BBH events between the $\Lambda$CDM model and the RHCT model. The solid black line and grey shaded region indicate the mean and 1-$\sigma$ uncertainty, respectively. The horizontal line indicates the threshold of ln(BF)=8 for confident model selection. A ln(BF) greater than 8 can be achieved with $\gtrsim 2100$ BBH detections, which corresponds to about two weeks of observations from ET.}
 \label{fig:BFL}
\end{figure}

Third-generation detectors like Einstein Telescope and Cosmic explorer will enable vast improvements in cosmological model selection. To investigate the typical sensitivity, we calculate the number of events required to distinguish the $\Lambda$CDM model from the RHCT model.

We create simulated data by taking a fixed number of random draws from a population of BBHs simulated using the $\Lambda$CDM model with $H_0=70 \,\si{km \ s^{-1}Mpc^{-1}}$ and $\Omega_m=0.3$. We then repeat the hyperparameter inference of Section~\ref{sec:hyperpe} and calculate the evidence assuming either a $\Lambda$CDM and RHCT cosmological model. Varying the number of random draws from the simulated data set, in Figure~\ref{fig:BFL} we plot the the Bayes factor for the $\Lambda$CDM vs. the RHCT model: a Bayes factor greater than unity indicates support for the $\Lambda$CDM model.

Figure \ref{fig:BFL} demonstrates that with just 2100 events, achievable in less than two weeks of observations of ET, the $\Lambda$CDM model can be distinguished from the RHCT model with a ln(BF) of 8.

\subsection{Discussion}
\label{sec:discuss}

We assume that the BBH mass spectrum does not evolve over cosmic time. Population synthesis studies have shown that the masses of BBHs are dependent on the metallicity of their progenitor stars \citep{Belczynski10max,Giacobbo18metal,safarzadeh2019impact}.
As the metallicity of star formation activities depends on redshift, it is possible that the BBH mass distribution evolves with redshift \citep[e.g.,][]{neijssel2019effect}.
However, this should not significantly affect our ability to perform cosmological inference as long as the key feature in the BBH mass spectrum remains stable over redshift.

Our analysis is based on the BBH population properties of GWTC-2; \citet{GWTC-2pop} found support for either a break or a bump at $\sim 40\, \si{M_{\odot}}$ on top of the power-law distribution for $m_1$.
At the moment, it remains unclear what is causing the feature at $\sim 40\, \si{M_{\odot}}$ seen in GWTC-2.
We conjecture it is related to PISN, but it may be due to other stellar processes or it could arise from the dynamics of assembly. If it is related to the PISN, the location and width of the PISN mass gap are robustly determined by nuclear physics \citep{Farmer19gap,Farmer20}.
Specifically, \citet{Farmer19gap} showed that the lower edge of the PISN mass gap is robust against variations in the metallicity, which implies that the PISN feature of the BBH mass spectrum should be relatively stable over redshift.
We expect that our understanding of the BBH mass function will be constantly improved as more events are detected.
If it turns out that the feature at $\sim 40\, \si{M_{\odot}}$ is not stable, it may be difficult to use it for precision cosmology.

\section{Conclusion}
\label{sec:conclusions}
Gravitational-wave astronomy provides a completely new way of studying the expansion history of our Universe.
In this paper we investigate how the stellar-mass black hole mass distribution, which contains a unique feature due to the pair instability supernova process, can be used to measure the Hubble constant and matter density without using electromagnetic counterparts or galaxy catalogues.
We show that for a third-generation detector like ET, one year operation with typically $10^5$ BBH detections will enable the Hubble constant to be measured to $\lesssim 1\%$, provided that the black hole mass distribution, cosmic star formation rate and binary merger delay time distribution is known \textit{a priori}.
Furthermore, as a proof of principle, we demonstrate that the alternative RHCT cosmology model can be distinguished from the standard $\Lambda$CDM model with merely two weeks of observations ($\sim 10^{4}$ BBH detections). This shows that gravitational-wave observations in the ET era can be a powerful tool for cosmological model selection.

We have developed a framework that allows simultaneous inference of cosmological parameters, the black hole mass distribution, the cosmic star formation rate, and the binary merger delay time distribution.
We find that the marginalization over current uncertainties in these processes reduces the $H_0$ measurement precision by nearly an order of magnitude.
This uncertainty is dominated by our incomplete understanding in the black hole mass distribution, which is likely to be overcome once $\gtrsim 10^3$ BBH detections are obtained with advanced detectors---assuming that the mass distribution does not evolve over cosmic time.

We expect our understanding of the BBH mass spectrum to be further refined with  $\mathcal{O}$(hundreds) of observations in the next few years.
These observations will allow us to better understand the physical origins of the features in the mass spectrum and potentially to establish the existence of subpopulations from dynamical and field BBH mergers. In this regard, future research is required to fully evaluate the prospect of cosmological inference using our approach.

\acknowledgements
This work was supported by the National Natural Science Foundation of China under Grants Nos. 11633001 and 11920101003 and also the Australian Research Council (ARC) Future Fellowship FT150100281 and Centre of Excellence CE170100004.
We thank Xilong Fan, Rory Smith, Francisco Hernandez, Colm Talbot, Boris Goncharov for useful discussions on this work.

\bibliography{references.bib}

\begin{figure}[htp]
 \centering
 \includegraphics[width=3.1in]{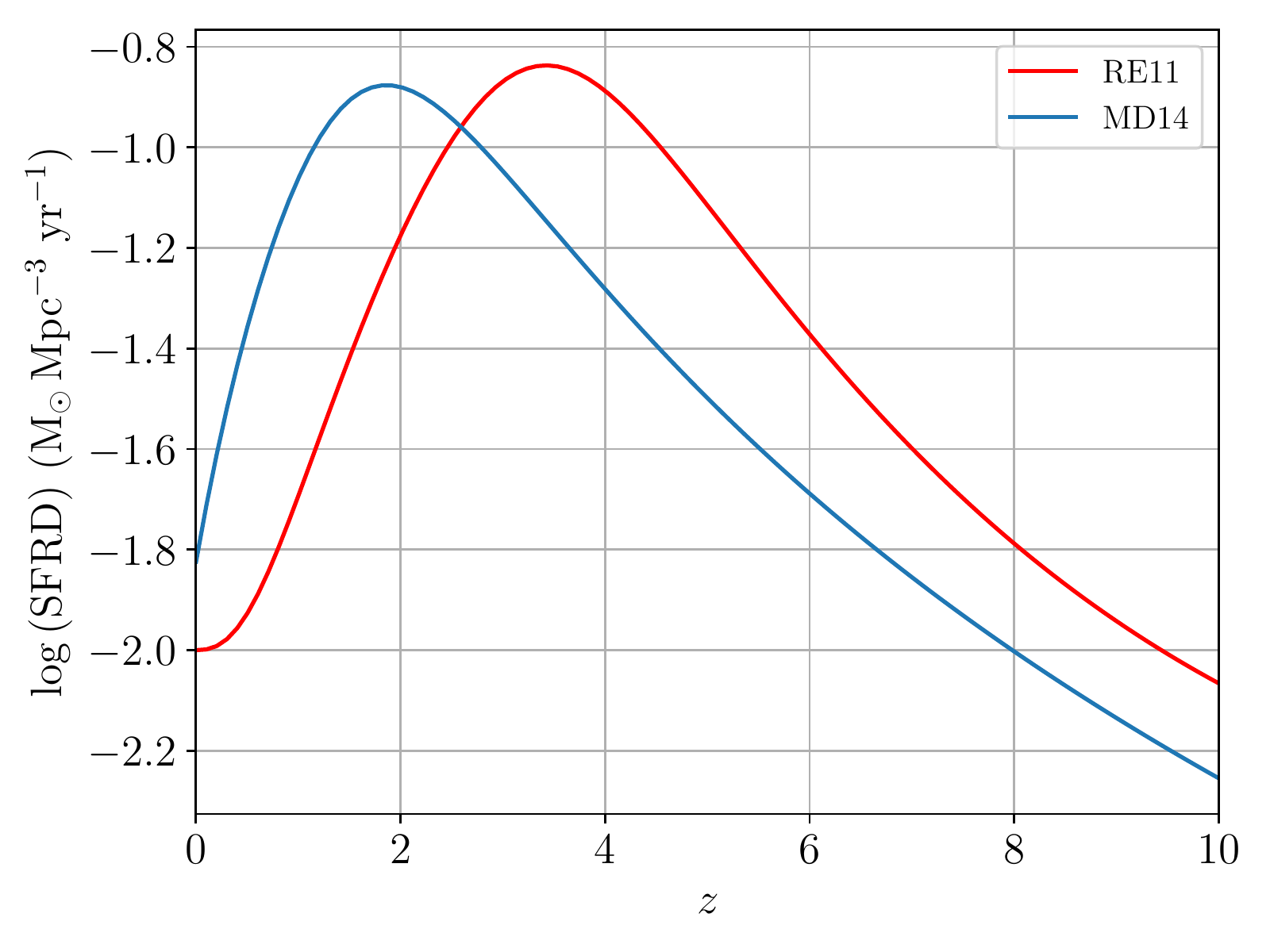}
 \includegraphics[width=2.45in]{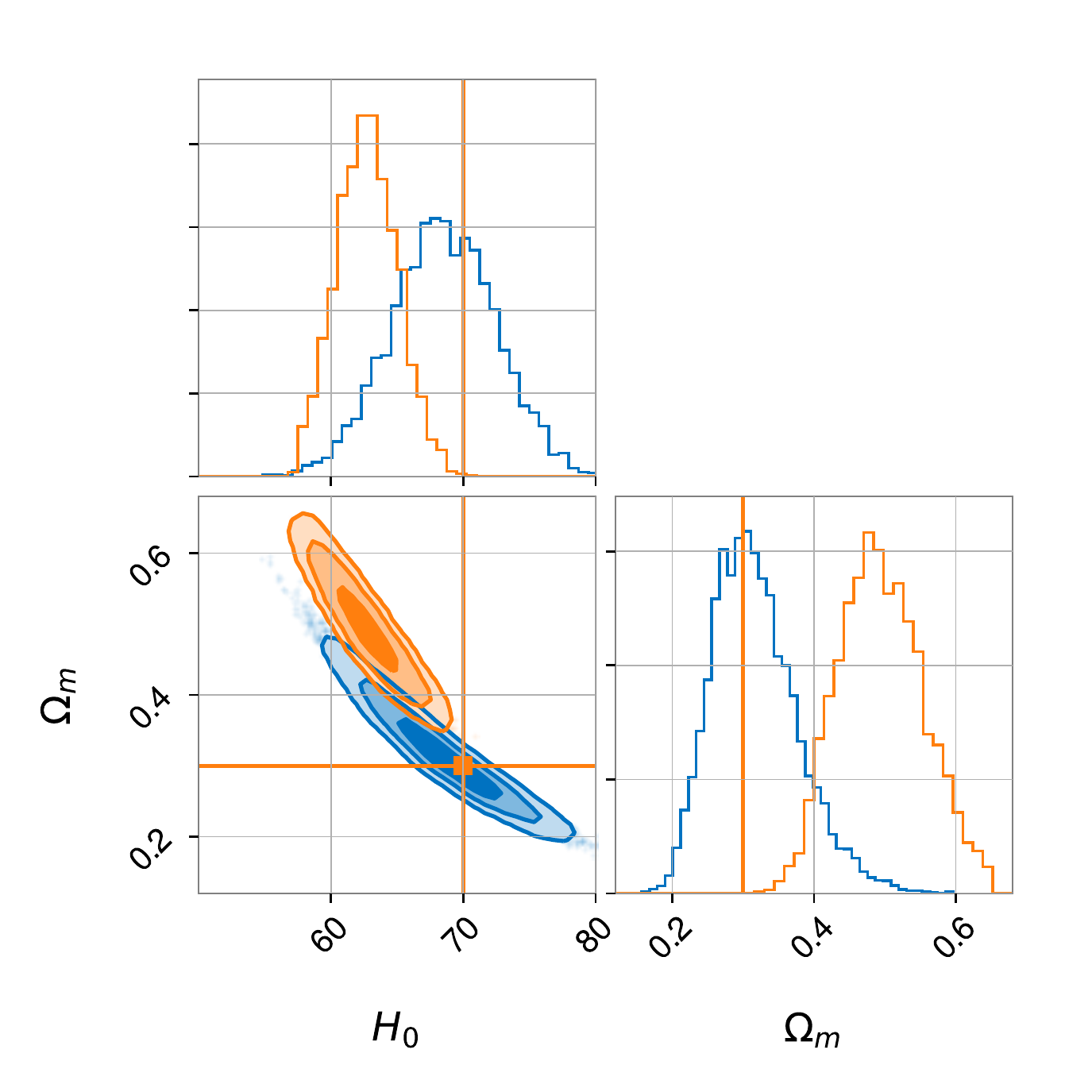}
 \caption{Left: The best-fit cosmic star formation rate density (SFRD) from \citet{madau2014cosmic} (MD14) and \citet{2012ApJ...744...95R} (RE11). Right: posterior distribution of $(H_{0}, \Omega_{m})$ derived using MD14 (blue) and RE11 (orange), while the MD14 model is used to generate injections.}
 \label{fig:sfrd}
\end{figure}

\appendix
\label{sec:append}
To demonstrate the effect of unknown cosmic star formation rate in cosmological inference using our approach, we consider two different models, MD14 \citep{madau2014cosmic} and RE11 \citep{2012ApJ...744...95R}, as shown in the left panel of Figure \ref{fig:sfrd}.
On the right panel of Figure \ref{fig:sfrd}, we show the posterior distribution of $(H_0,\,\Omega_m)$ using these two models while the true underlying model is MD14.
One can see that the estimates of  $(H_0,\,\Omega_m)$ are biased (orange contours) if an incorrect star formation rate model is used.

\begin{figure}[htp]
 \centering
 \includegraphics[width=7.2in]{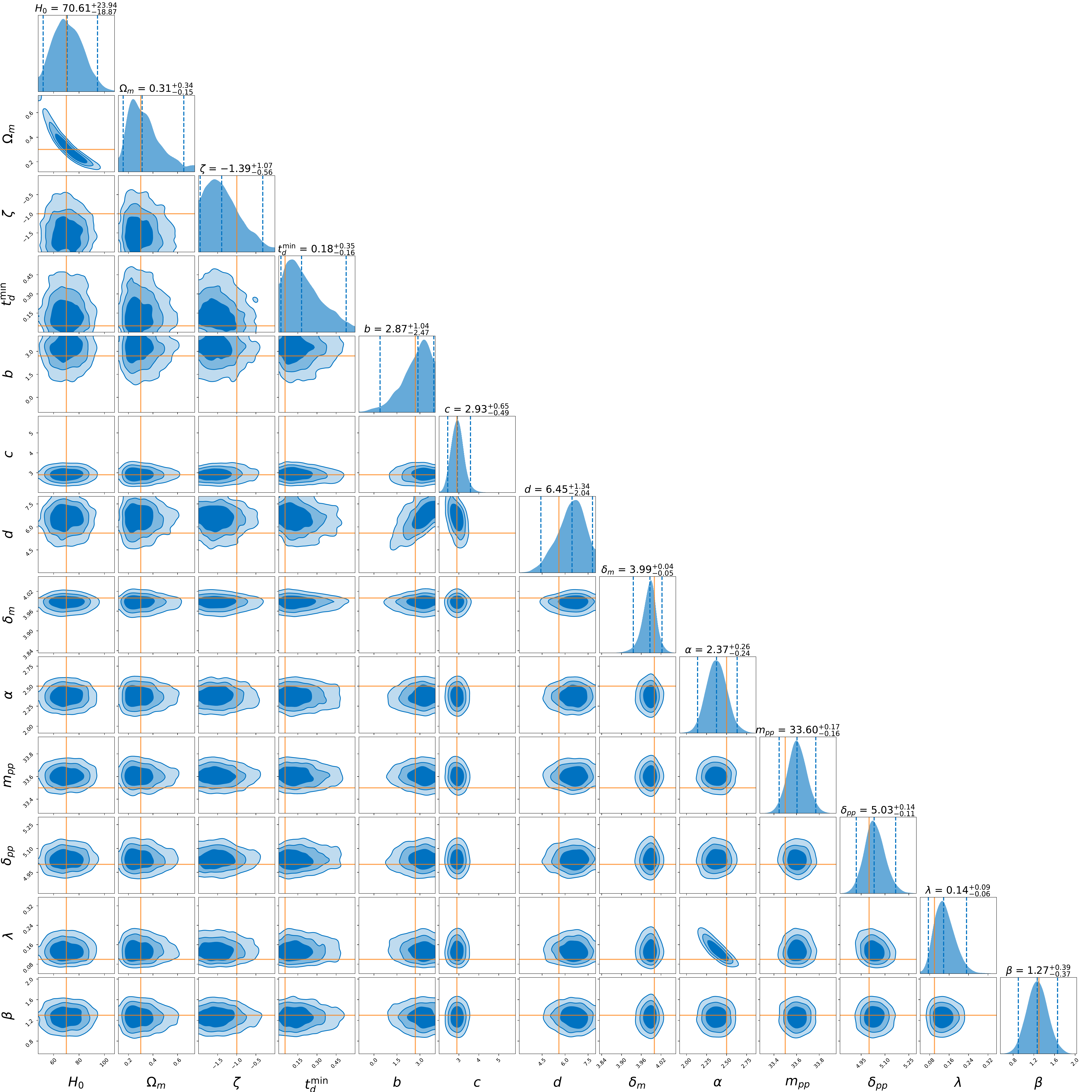}
 \caption{Posterior distributions for hyperparameters $(H_0,\, \Omega_m,\, \zeta,\, t_{d}^{\rm{min}},\,b,\,c,\,d,\,\delta_m,\,\alpha,\,m_{\rm{pp}},\,\delta_{\rm{pp}},\,\lambda,\,\beta)$ with $10^{3}$ zero-error BBH injections. }
 \label{fig:8p}
\end{figure}

Figure \ref{fig:8p} shows the posterior distributions of all hyperparameters derived using $10^{3}$ zero-error injections. These parameters include cosmology parameters $(H_{0}, \Omega_{m})$, cosmic star formation rate density parameters $(b,\,c,\,d)$, delay time distribution parameters $(\zeta,\,t_{d}^{\rm{min}})$, and black hole mass distribution parameters $(\delta_m,\, \alpha,\,m_{\rm{pp}},\,\delta_{\rm{pp}},\,\lambda,\,\beta)$ while fixing $m_{\rm{min}}=5\, {\rm{M_{\odot}}}, \, m_{\rm{max}}=65\, \rm{M_{\odot}}$.
We adopt uniform priors for all hyperparameters within ranges specified as follows: $H_0 \in [40,105] \, \si{ km \ s^{-1}Mpc^{-1}}$, $\Omega_m \in [0,0.75]$, $\zeta \in [-2,0],\, t_{d}^{\rm{min}} \in [0.001,1] \, \si{Gyr},\,b \in [0,5],\,c \in [0,5],\,d \in [0,10],\,\delta_m \in [0,10] \, \si{M_{\odot}},\,\alpha \in [-4,12],\,m_{\rm{pp}} \in [20,50]\, \si{M_{\odot}},\,\delta_{\rm{pp}} \in [0.4,10]\, \si{
M_{\odot}},\,\lambda \in [0,1], \rm{and} \, \beta \in [-4,12]$.
Hyper priors for the BBH mass model are the same as listed in Table 5 in \citet{GWTC-2pop}.

\end{document}